\documentclass{aa}  
\usepackage{graphicx}
\usepackage{txfonts}
\def\h2{\ifmmode {\mbox H$_2$}\else H$_2$\fi\xspace}

\begin{document}

   \title{ALMA reveals starburst-like interstellar medium conditions in a compact star-forming galaxy at $z\sim 2$
     using [CI] and CO}
   \author{Gerg\"o Popping
          \inst{1}, 
          Roberto Decarli\inst{2}, Allison W. S. Man\inst{1}, Erica J. 
          Nelson\inst{3}, Matthieu B\'ethermin\inst{1,4}, 
          Carlos De Breuck\inst{1}, Vincenzo Mainieri\inst{1}, Pieter G.
          van Dokkum\inst{5}, Bitten
          Gullberg\inst{6}, Eelco van Kampen\inst{1}, Marco
          Spaans\inst{7}, Scott C. Trager\inst{7} 
          }          
   \institute{European Southern Observatory, Karl-Schwarzschild-Strasse 2, 85748, Garching, Germany
              \email{gpopping@eso.org}
         \and
             Max-Planck Institut f\"ur Astronomie, K\"onigstuhl 17, D- 69117, Heidelberg, Germany
             \and
             Max-Planck-Institut f\"ur Extraterrestrische Physik
             (MPE), Giessenbachstr., D-85748 Garching
           \and Aix Marseille Univ, CNRS, LAM, Laboratoire d'Astrophysique de Marseille, Marseille, France
             \and Astronomy Department, Yale University, New Haven, CT
             06511, USA
             \and 
             Centre for Extragalactic Astronomy, Department of Physics, Durham University, South Road, Durham DH1 3LE, UK
             \and
             Kapteyn Astronomical Institute, University of Groningen, Postbus 800, NL-9700 AV Groningen, the Netherlands\\
}
\titlerunning{Starburst-like interstellar medium conditions in a compact star-forming galaxy at $z\sim 2$}
\authorrunning{Gerg\"o Popping et al.}

  \abstract
  {We present ALMA detections of  the [CI] 1--0, CO J$=$3--2, and CO
     J$=$4--3 emission lines, as well as the ALMA band 4 continuum for a compact
     star-forming galaxy (cSFG) at $z=2.225$, 3D-HST GS30274.  As is typical for cSFGs, this galaxy has a stellar mass of 
     $1.89 \pm 0.47\,\times 10^{11}\,\rm{M}_\odot$, 
     with a star formation rate of $214\pm44\,\rm{M}_\odot\,\rm{yr}^{-1}$
     putting it on the star-forming `main-sequence', 
     but with an H-band effective radius of 2.5\,kpc, making it much 
     smaller than the bulk of `main-sequence' star-forming galaxies.
     The intensity ratio
     of the line detections yield an ISM density ($\sim
     6 \times 10^{4}\,\rm{cm}^{-3}$) and a UV-radiation field ($\sim
     2 \times 10^4\,\rm{G}_0$), similar to the values in local
     starburst and ultra-luminous infrared galaxy 
     environments. A starburst phase is consistent with the short depletion
     times ($t_{\rm H2, dep} \leq 140$ Myr) we find in 3D-HST GS30274
     using three different proxies for the \h2 mass ([CI], CO, dust mass). This depletion time is significantly shorter than in
     more extended SFGs with similar stellar masses and
     SFRs. Moreover, the gas fraction of 3D-HST GS30274 is smaller than
     typically found in extended galaxies. 
     We measure the CO and [CI] kinematics and find a FWHM line width 
     of $\sim 750 \pm 41 $ km s$^{-1}$. The CO and [CI] FWHM are consistent with
     a previously measured H$\alpha$ FWHM for this source. The line
     widths are consistent with gravitational motions, 
       suggesting we are seeing a compact molecular gas reservoir. A previous merger event, as suggested by the
     asymmetric light profile, may be responsible for the compact distribution of gas and
     has triggered a central starburst event. This event gives rise to the
     starburst-like ISM properties and short depletion times in 3D-HST GS30274. The
     centrally located and efficient star formation is quickly building up a
     dense core of stars, responsible for the compact distribution
     of stellar light in 3D-HST GS30274.
}

   \maketitle
%

\section{Introduction}
Recently, significant progress has been made in understanding the formation 
of the dense cores of massive galaxies. These forming cores have been 
discovered in the form of a population of massive 
($M_* > 10^{10}\,\rm{M}_\odot$), compact, star-forming galaxies (cSFGs) at
$z\geq2$
\citep{Barro2013,Barro2014a,Nelson2014,Williams2014,vanDokkum2015}.
cSFGs are in a dusty star-forming phase 
\citep{Nelson2014,Barro2014a,vanDokkum2015} with SFRs similar
to or slightly lower than in star-forming galaxies on the
main-sequence of star formation. cSFGs galaxies have similar morphological properties as compact quiescent
galaxies, including high S\'ersic indices, and centrally concentrated luminosity profiles. A comparison in the
evolution of their respective number densities suggested that cSFGs
are the direct progenitors of compact quenched objects
\citep{Barro2013,vanDokkum2015}. This hypothesis was supported by NIR
spectroscopy of a sample of cSFGs which revealed that their
H$\alpha$ and [OIII] emission line widths are similar to the observed stellar
velocity dispersion of compact quiescent galaxies
\citep{Nelson2014,Barro2014b}.

\citet{Barro2014b} found that cSFGs have integrated
velocity dispersions with full-width-half-maximum (FWHM) velocities as high
as 600 km s$^{-1}$. \citet{vanDokkum2015} argued that the
broad H$\alpha$ line widths are driven by reservoirs of
centrally-located rapidly-rotating gas. High spatial-resolution ALMA
observations of the sub-mm continuum in cSFGs show that these
galaxies have very compact dust emission, up to more than a factor of
two smaller than the stellar emission
\citep{Barro2016,Tadaki2016}. From the observed dust emission,
\citet{Barro2016} and \citet{Tadaki2016} measured gas depletion times
of only a few hundred Myr, similar to their estimated quenching
timescales. The compact stellar, 
and dust morphologies are likely
the result of dissipational processes, which in turn increase the
velocity dispersions \citep{Barro2013,Barro2014a,vanDokkum2015}. This
scenario for the compact distribution of matter in cSFGs is supported by
theoretical models. These suggest that disc instabilities or gas-rich merger
events can indeed drive the compaction of gas and induce a nuclear starburst
event that quickly reduces the galaxy half-light radius \citep{Barnes1991,Mihos1994,Wellons2015,Zolotov2015}. The nuclear
starburst will quickly build a compact stellar centre, possibly drive
an outflow, and transform larger, star-forming galaxies into compact
quiescent galaxies over a short time
\citep{Zolotov2015}. \citet{Genzel2014} observed the signatures
of strong nuclear outflows in the ionized gas kinematics of massive
galaxies, some of which can be classified as cSFGs.

The properties of the atomic and molecular gas are a key
  piece of information missing in this scenario. Because the shut down of 
   star formation is thought to be imminent in cSFGs \citep{vanDokkum2015}, 
   a measurement of their gas mass and depletion time is essential. 
  If cSFGs have a strong ongoing nuclear starburst, it is to be expected that the 
  molecular gas has a short depletion time. The gas should furthermore have high densities with a strong impinging 
  radiation field from young stars.  The spatial distribution and kinematics of ionized 
   gas suggests that cSFGs have rotating disks of ionized gas that 
   are a factor of two more extended than the stellar distribution 
   \citep{vanDokkum2015}. A key test of whether we are actually observing
   the building of the dense cores of massive galaxies through in situ
   star formation is a measurement of a high line width in molecular
   gas \citep[e.g.,][]{Tacconi2006,Tacconi2008}. Sub-mm line information probing the molecular gas content, the
molecular gas conditions (density, radiation field), and the molecular gas kinematics
are thus crucial to understand the nature of cSFGs.

\citet{Barro2016} used ALMA band 7 continuum imaging to probe the gas
depletion time of six cSFGs and found that they are typically
shorter than in extended main-sequence galaxies with similar stellar
masses, SFRs, and redshift. Different groups have observed the gas content of extended main-sequence
galaxies at $z>0$ either through their $^{12}$CO (hereafter
CO) emission or the dust continuum
\citep[e.g.,][]{Aravena2010,Daddi2010,Tacconi2010, Geach2011, Aravena2012,Magdis2012a,Magdis2012b,
  Bauermeister2013, Saintonge2013, Tacconi2013,Tan2013,Santini2014,
  Bethermin2015,Daddi2015,Genzel2015,
  Berta2016,Decarli2016,Scoville2016}. Both approaches (CO and dust) rely on uncertain conversion factors between the observed
luminosity and an estimated gas mass. Independent measures
of the gas mass in galaxies are therefore necessary to overcome the
systematic uncertainty in these
conversion factors and constrain galaxy gas masses. This is
especially relevant in cosmic-ray or X-ray dominated environments
such as starbursts and AGN hosting galaxies
\citep{Bisbas2015,Glover2016}, the environments one can expect to
exist in cSFGs.

Atomic carbon ([CI]) can be used to measure the physical properties of
the ISM in addition to
the conventional methods described above. 
The optically thin [CI] is a good
tracer of the surface of molecular
clouds. The
optically thick CO traces the gas in cores shielded from UV
radiation. Cosmic ray ionization allows for the
co-existence of CO and [CI], for instance in starburst
environments. [CI] emission can thus especially in these environments be used as a
good tracer of the \h2
mass in galaxies
\citep{Weiss2005,Papadopoulos2004}. \citet{Bisbas2015} indeed showed that
under the influence of a strong cosmic-ray field [CI] is a better
tracer of the \h2 mass than CO. Independently, \citet{Glover2016} also
found that [CI] is a better tracer of \h2 in cosmic-ray
dominated regions.  Additionally, the line ratios between [CI] and CO emission
can be used to constrain the UV radiation field and density of the ISM
\citep[e.g.,][]{Meijerink2007,Alaghband2013,Bothwell2016}. Combining observations of the carbon
fine-structure lines with CO is therefore a great
strategy to build a more complete picture of the ISM characteristics
(\h2 mass, gas density, and UV radiation field)
and study the mode of star formation in cSFGs. 
%
The CO and [CI] emission line
detections can simultaneously be used to study the dynamics of the atomic and
molecular gas in cSFGs.

In this paper we present Atacama Large sub/millimeter array (ALMA)
detections of the CO J$=$3--2, CO J$=$4--3, and the [CI]
$^3$P$_1$--$^3$P$_0$ (hereafter [CI] 1--0) emission lines and band 4
continuum in a cSFG at $z=2.225$, 3D-HST GS30274.  We study CO and [CI]
kinematics, the density and UV radiation field of the ISM, and the
global gas properties such as gas fraction and gas depletion time. In Section \ref{sec:observations} we
present the selection of 3D-HST GS30274 and the observations. The line and
continuum detection of 3D-HST GS30274 and their analysis are presented in Section
\ref{sec:results}.  We discuss our findings in Section
\ref{sec:discussion} and present our conclusion in Section \ref{sec:summary}. Throughout this paper we
assume a Chabrier initial mass function \citep{Chabrier2003}. All
presented gas masses are pure molecular hydrogen-masses and do not include a
correction for Helium unless stated otherwise.

\begin{figure}[t]
\centering
\includegraphics[width=\linewidth]{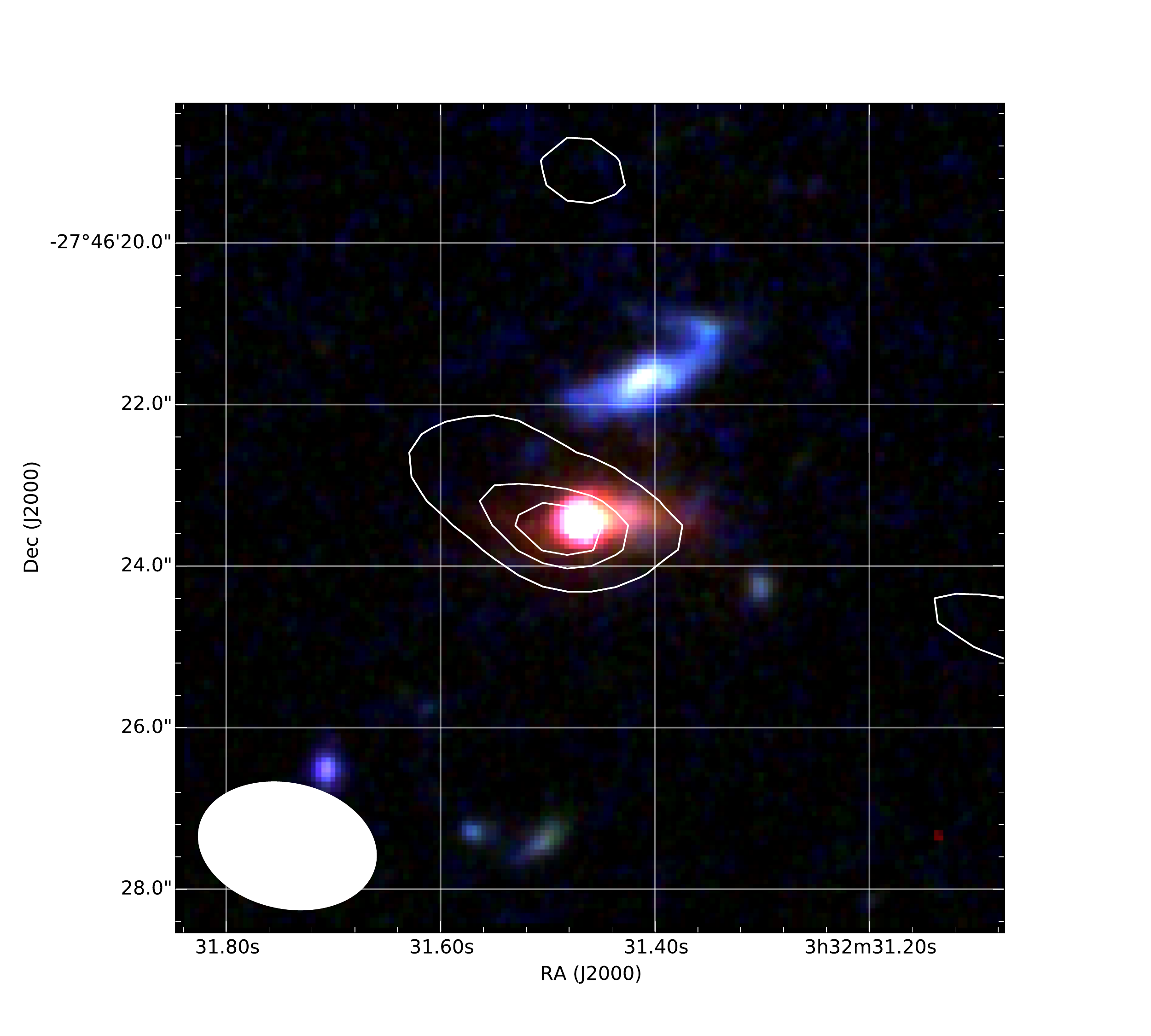}
\centering
 \caption{\label{fig:H_band}{A Hubble Space Telescope  F160W/F105W/F435W  RGB image of 3D-HST
     GS30274. The images were  obtained as a part of the Cosmic Assembly
Near-infrared Deep Extragalactic Legacy Survey
\citep[CANDELS,][]{Grogin2011,Koekemoer2011}. The source to the north of our target is a lower-redshift foreground
     galaxy. The ALMA band 4 continuum detection of 3D-HST
     GS30274 is shown as contours (2, 3, 4, and 5 sigma). The HST RGB shows an asymmetric light-profile to the west of
     the core in 3D-HST GS30274, indicative of a recent merger event.}}
\end{figure}

\begin{figure}[t]
\centering
\includegraphics[width=\linewidth]{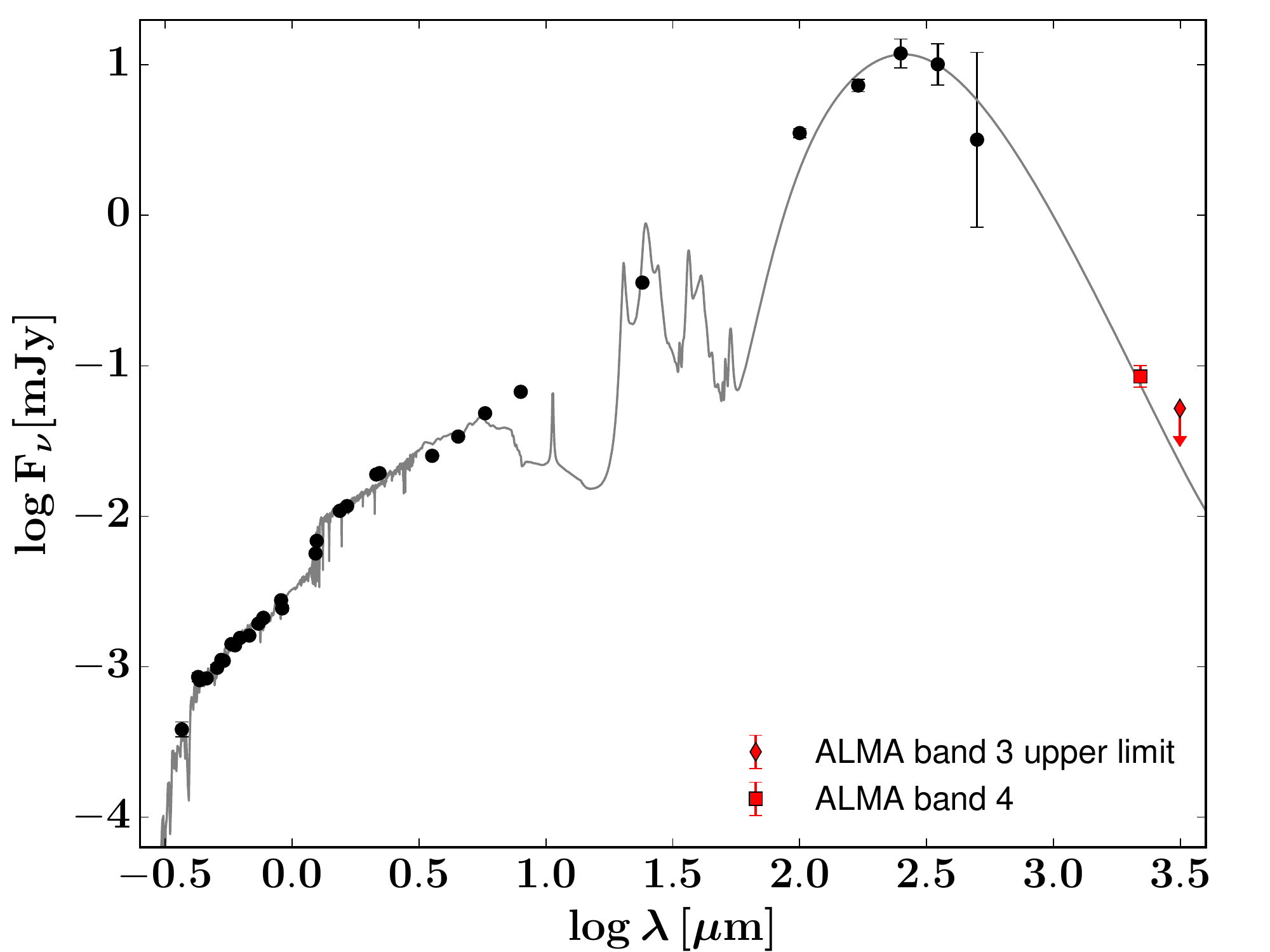}
\centering
 \caption{\label{fig:SED}{ Multi-wavelength spectral
     energy distribution of 3D-HST GS30274. The best fit from the \texttt{MAGPHYS} analysis is plotted
     as a line. The optical, NIR,
     \emph{Herschel} measurements are shown as bold circles. The ALMA
     band 4 continuum detection is presented as a red square. The red
     diamond represents the three sigma upper limit on the ALMA band 3 continuum.}}
\end{figure}

\begin{figure*}[t]
\centering
\includegraphics[width=\linewidth]{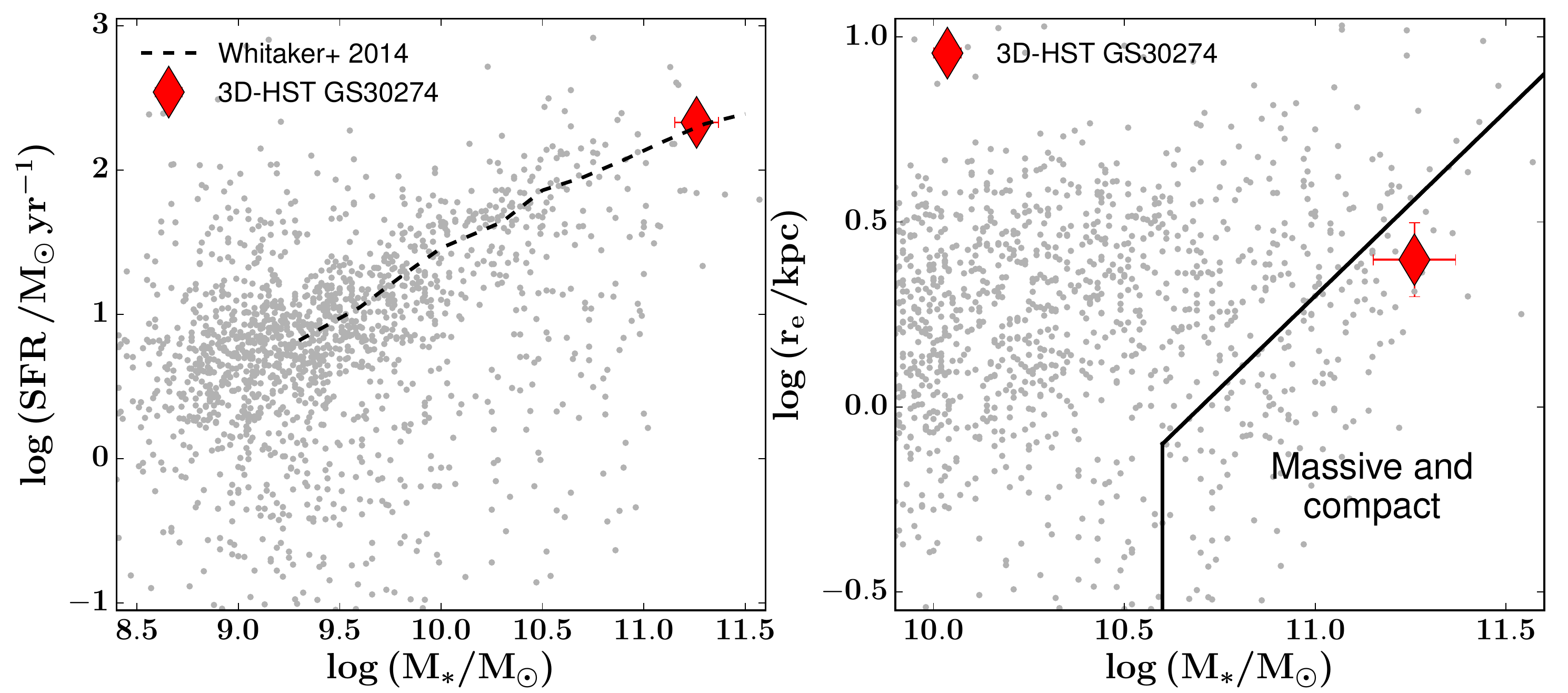}
\centering
 \caption{\label{fig:MS} Left: The location of 3D-HST GS30274 in the stellar
     mass vs. SFR plane of galaxies. The grey dots show galaxies
     taken from all five 3D-HST/CANDELS fields \citep{Momcheva2016} in
     the redshift range 2.0<z<2.5. The black line show a fit to the
     these galaxies from \citet{Whitaker2014}. The
source in this study, 3D-HST GS30274, shown as a red diamond, lies on top
of the main-sequence of galaxies at these redshifts, at the high-mass
end.  Right: The size-mass relation of galaxies with redshifts
$2.0<z<2.5$ from the five 3D-HST/CANDELS fields
\citep{Momcheva2016,vanderWel2014}. The solid lines define the \citet{vanDokkum2015} selection
criteria for compact, massive galaxies: $\log{(M_*/\rm{M}_\odot)} >
10.5$ and $\log{(r_e/\rm{kpc})} < \log{(M_*/\rm{M}_\odot)} - 10.7$. 3D-HST GS30274 is shown as a red
diamond and falls within the area for compact, massive galaxies.}
\end{figure*}

\begin{table*}[]
\centering
\caption[]{\label{tab:globproperties}{{Global properties of 3D-HST GS30274. }}}
\begin{tabular}{cccccccccc}
\hline \noalign {\smallskip}
\hline \noalign {\smallskip}
RA & Dec & z & $M_{*}$ & $\rm{SFR} $ & $r_{e}$ & $M_{\rm dust}$ &  T$_{\rm dust}$
  & 
                                                                      $M_{\rm
                                                                      [CI]}$
                                                                      &
                                                                        $L_{IR}$\\
deg &deg&&$10^{11}\,\rm{M}_\odot$&$\rm{M}_\odot\,\rm{yr}^{-1}$
                       
                                     &kpc&$10^{8}\,\rm{M}_\odot$&K &$10^6\,
                                                              \rm{M}_\odot$
                                                      & 

$10^{12}\,\rm{L}_\odot$\\
\hline \noalign {\smallskip}
53.13108 & -27.77311 & 2.225  & $1.89\pm0.47$ &$214\pm44$&
                                                           $2.5\pm0.1$&$2.5\pm0.6$
                                                                        &$45
                                                                        \pm
                                                                        5$&
                                                                          $5.4
                                                                        \pm 1.4$&$2.3\pm0.6$\\
\hline \noalign {\smallskip}
\end{tabular}
\end{table*}

\section{Observations and Data reduction}
\label{sec:observations}
\subsection{Target selection}
To select a cSFG a preliminary sample was drawn from the GOODS-South region in the 3D-HST 
catalogue \citep{Skelton2014} that includes photometry from the U-band
to the {\emph Spitzer} 8 $\mu$m band. 
Using a BzK selection technique
\citep{Daddi2004} we
selected galaxies that fall on the main-sequence of star-formation \citep[][]{Noeske2007,Daddi2008,Whitaker2014}.  We
then requested the galaxies to have 1) A secure spectroscopic redshift between 2.0 and
  2.5, to assure that the CO J$=$4--3 and the [CI] 1--0 line could be
  observed simultaneously within the same tuning; 2) A counterpart in the blind fields from
  \emph{Herschel}/PACS \citep{Magnelli2013} and SPIRE
  \citep{Roseboom2010};

From the galaxies in the GOODS-South region from the 3D-HST survey that
meet these criteria, we select the most massive cSFG following the
classification by \citet[$\log{(M_*/\rm{M}_\odot)} >
10.5$ and $\log{(r_e/\rm{kpc})} < \log{(M_*/\rm{M}_\odot)} -
10.7$]{vanDokkum2015}: 3D-HST GS30274. This source
has an effective radius of 2.5 $\pm$ 0.1 kpc
\citep[flag for the fit equals 2, which marks it as a bad fit]{vanderWel2014}. The S\'ersic index of this galaxy is not well
defined. The poor flag for the fitting and the undefined S\'ersic index are the result of an asymmetric
stellar light profile to the west of the centre of 3D-HST GS30274, suggestive of a merger remnant (Figure
\ref{fig:H_band}). 3D-HST GS30274 is a hard X-ray source
with a luminosity of $1.15\times 10^{43}\,\rm{erg}\,\rm{s}^{-1}$,
classified to have an active galactic nucleus
\citep[AGN;][]{Xue2011,Luo2016}. Line ratios
of the H$\alpha$, [OI], [OIII], [SII], and [NII] emission lines also
suggest 3D-HST GS30274 has an AGN
\citep{vanDokkum2005,Genzel2014}. Moreover, this object is detected in the 1.4 GHz
Very Large Array catalogue of the Extended Chandra Deep Field South
with a flux of $84.2\pm  6.1\,\mu$Jy
\citep{Bonzini2012}. \citet{Genzel2014} attributed the broad H$\alpha$
 profile in 3D-HST GS30274 to be outflow-driven, though \citet{vanDokkum2015}
argues that a large fraction of the broad line width can be explained
by compact quickly rotating ionized gas.

The stellar mass and SFR for this galaxy were calculated by fitting
the spectral energy distribution (SED) using \texttt{MAGPHYS} \citep{daCunha2008,daCunha2015}, including photometry from
\citet{Skelton2014} as well as the \emph{Herschel} data presented in
\citet{Magnelli2013} and \citet{Roseboom2010}. Including also the ALMA
band 4 continuum (see Section 3) we find a stellar mass and SFR of $1.89 \pm 0.47\,\times 10^{11}\,\rm{M}_\odot$ and
$214\pm44\,\rm{M}_\odot\,\rm{yr}^{-1}$, respectively. We present the photometry used for
the SED fitting in Table \ref{tab:photometry}. The
  spectral energy distribution (SED) and the fit to it is
presented in Figure \ref{fig:SED}. The global properties of 3D-HST GS30274 that we obtained from the SED fitting are presented in Table \ref{tab:globproperties}.

The location of 3D-HST GS30274 on the stellar mass -- SFR relation
and stellar mass -- size relation of galaxies is shown in Figure
\ref{fig:MS}. Its compact size
and its location on the main-sequence meet the criteria set by
\citet{Barro2013} and \citet{vanDokkum2015} to classify 3D-HST GS30274
as a typical cSFG.

\subsection{Observations}
The ALMA observations of 3D-HST GS30274 were carried out under project
2015.1.00228.S (PI: G. Popping). Band 3 observations of 3D-HST GS30274
were performed on January 12 2016 with a compact configuration
(C36-1; shortest and longest baselines were 15.1 and 331 meters, respectively
with an rms of 126.9 meters). The water vapour during the observations
was PWV $= 5$ mm. The integration time on 3D-HST GS30274 was 46
minutes. The data were calibrated using the standard ALMA pipeline
(\texttt{CASA} version 4.5.1). One spectral window was centered at
99.322 GHz with a bandwidth of 1.875 GHz covering 3840 channels, to
target the CO J$=$3--2 emission line. The other spectral windows were
centered at 95.182, 107.199, and 105.241 GHz with a bandwidth of
1.875 GHz covering 128 channels. These spectral windows were used to
observe the band 3 continuum of 3D-HST GS30274. Images were produced
using the \texttt{CLEAN} task in \texttt{CASA} \citep{McMullin2007},
with a natural weighting. This resulted in a synthetic beam size of
3.4 $\times$ 2.16 arcsec$^2$ with a position angle of 274 degrees. The typical rms noise level is 0.012 mJy/beam for the continuum and 0.16 mJy/beam/(100 km s$^{-1}$) for the CO J$=$ 3--2 line. 

The band 4 observations of 3D-HST GS30274 were performed on January 17
2016 and  January 21 2016 with the same configuration as for the band
3 observations. The water vapour during the observations was PWV
$=4.4$ mm on January the 17th and PWV $=2$ mm on January the 21st. The total integration time on 3D-HST GS30274 in band 4 was 89 minutes. 
The data were calibrated using the standard ALMA pipeline. One
spectral window was centered at 142.568 GHz with a bandwidth of 1.875
GHz covering 240 channels, to target the CO J$=$4--3 emission line. A
second spectral window was centered at 152.610 GHz with a bandwidth of
1.875 GHz covering 240 channels to target the [CI] 1--0 emission
line. Two additional spectral windows were placed at 140.693 and
154.551 GHz with a bandwidth of  1.875 GHz covering 128 channels. These
spectral windows were used to measure the band 4 continuum emission of 3D-HST GS30274. 
The imaging was done with the \texttt{CLEAN} task with natural
weighting and resulted in a synthetic beam size of 2.22 $\times$ 1.54
arcsec$^2$ with a position angle of 76 degrees. The typical rms noise level is 0.014 mJy/beam for the
continuum and 0.076 mJy/beam/(100 km s$^{-1}$) for the CO J$=$ 4--3 line and 0.08 mJy/beam/(100 km s$^{-1}$) for the [CI] 1--0 line. 

\begin{table}[]
\centering
\caption[]{\label{tab:submmproperties} The observed sub-mm properties of 3D-HST GS30274}
\begin{tabular}{cccc}
\hline \noalign {\smallskip}
\hline \noalign {\smallskip}
Feature & I$_{\rm line}$ & L'$_{\rm line}$ & S$_{660 \mu m}$ \\
& Jy km/s &  $10^9$ K km/s pc$^{2}$&  mJy\\
\hline \noalign {\smallskip}
CO (3--2) & $0.77 \pm 0.06$ &$21.0 \pm 1.6$&...\\
CO (4--3) & $1.08 \pm 0.04$ &$16.5\pm 0.5$ &...\\
$[$CI$]$ (1--0) & $0.32\pm 0.03$ &$4.3 \pm 0.4$ &...\\
660 $\mu \rm{m}^a$& ... & ...& $0.085 \pm 0.014$\\
\hline \noalign {\smallskip}
\end{tabular}
\begin{flushleft}{\small $^a$ restframe wavelength}\end{flushleft}
\end{table}

\begin{figure*}[t]
\centering
\includegraphics[width=\linewidth]{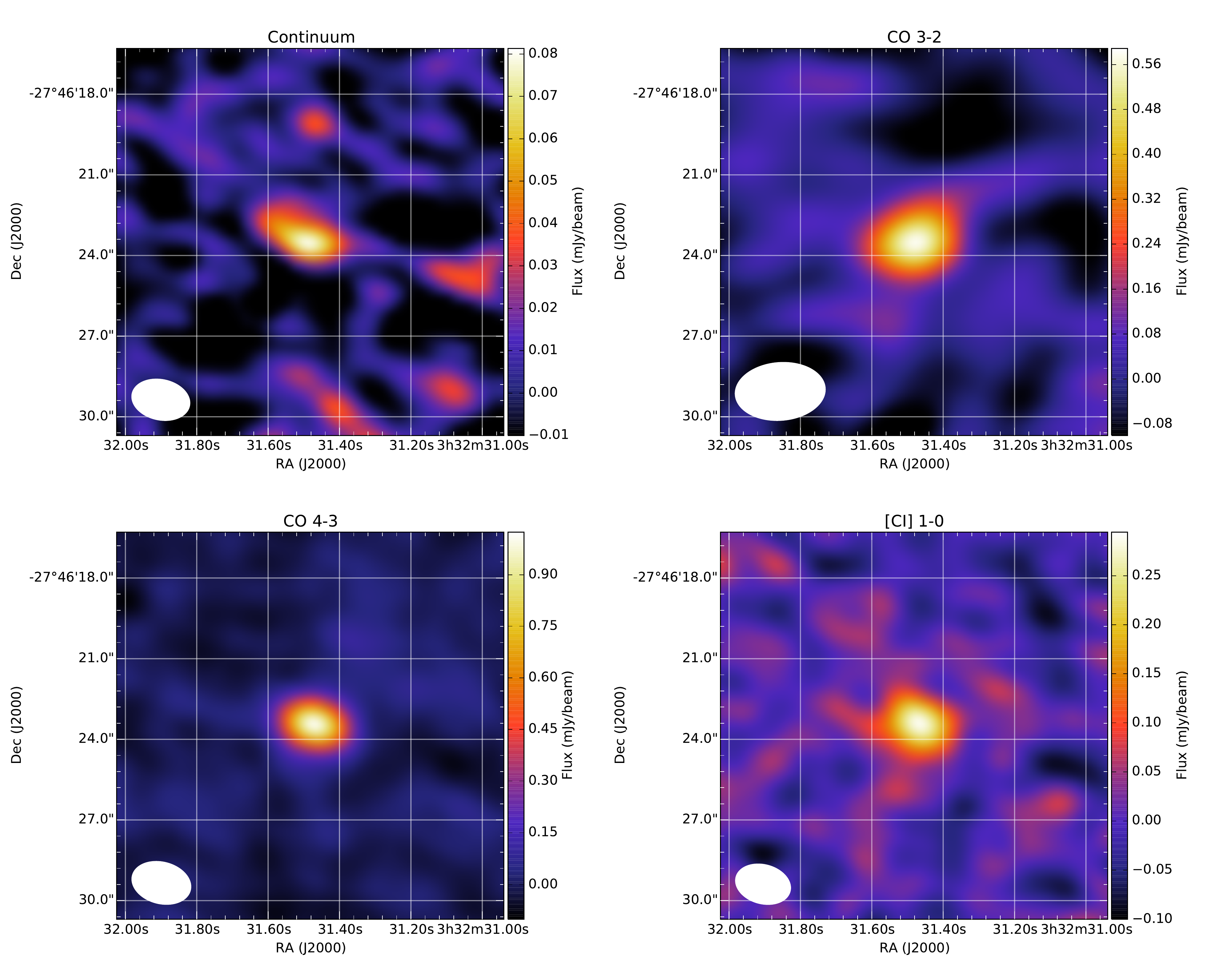}
\centering
 \caption{\label{fig:poststamp}{Moment zero maps of the band 4 continuum (top left), the
     CO J$=$3--2 line (top right), the CO J$=$4--3 line (bottom left),
     and the [CI] 1--0 line (bottom right). The size of the beam is indicated in the lower left
corner. The lines were integrated between -600 and +600 km s$^{-1}$
. All the lines and the underlying continuum in band 4 are clearly
detected.}}
\end{figure*}

\begin{figure*}[t]
\centering
\includegraphics[width=\linewidth]{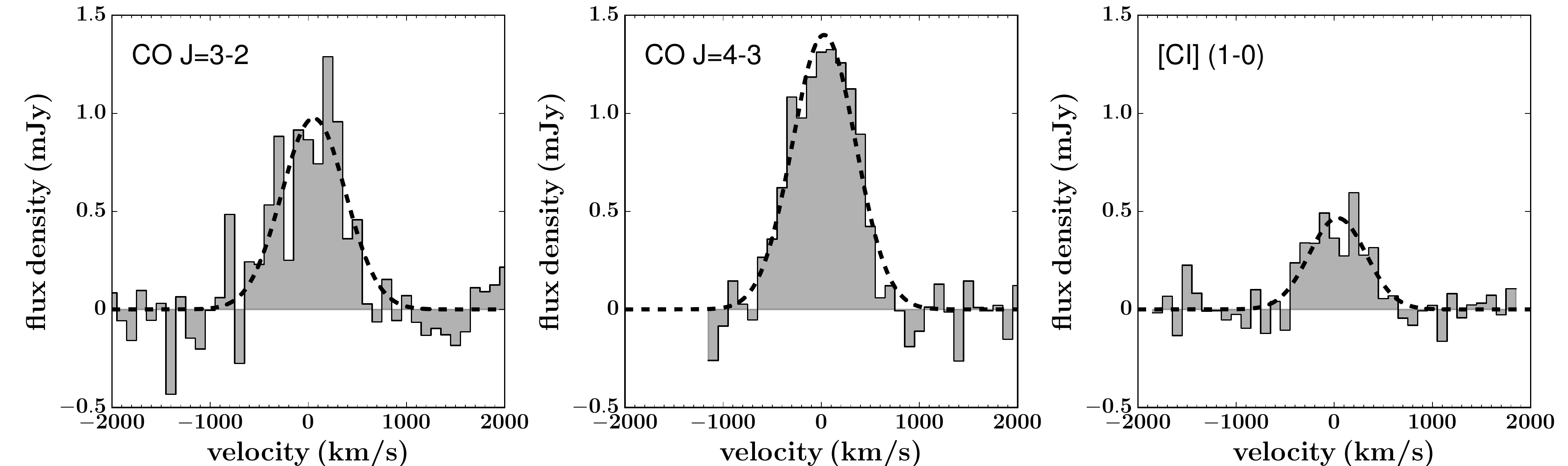}
\centering
 \caption{\label{fig:spectra}{ Integrated flux density of the CO J$=$3--2
  (left), CO J$=$4--3 (middle), and [CI] 1--0 (right) emission lines in
  3D-HST GS30274. The dashed line marks the gaussian fit to the density
  profile. All the lines are clearly detected and can be well fitted
  by a gaussian}}
\end{figure*}

\section{Results and Analysis}
\label{sec:results}
\subsection{Emission line and continuum detections}
We present moment zero maps of the CO J$=$3--2, J$=$4--3, and
[CI] 1--0 emission of 3D-HST GS30274 in Figure \ref{fig:poststamp}. We also
show the band 4 continuum at a rest-frame wavelength of 660 $\mu
m$.  We see clear detections of the targeted CO and [CI] emission lines and
the band 4 continuum. We detect the CO
J$=$3--2, J$=$4--3, and the [CI] 1--0 line a signa-to-noise ratio of
12.8, 27,
  and 10.7, respectively. The band 4 continuum at a
rest-frame wavelength of 660 $\mu m$ is detected with a
signal-to-noise ratio of 6.1. The band 3 continuum at a rest-frame wavelength of 976 $\mu
m$ is not detected, consistent with the flux intensity predicted by
the \texttt{MAGPHYS} SED fit (Figure \ref{fig:SED}). In none of the cases do we spatially resolve
the galaxy. This was to be expected, as its near-infrared half-light
radius \citep[0.53 arcsec][]{Skelton2014} is significantly smaller
than the minor axis of the primary beam obtained in the ALMA images
(2.4 arcsec in band 3 and 1.54 arcsec in band 4).

We show the flux density profile of the CO J$=$3--2, J$=$4--3, and
[CI] 1--0 detections in Figure \ref{fig:spectra}. We calculate the CO and [CI] line luminosities using the following
relation \citep{Solomon2005}
\begin{equation}
\rm{L}'\,(\rm{K}\,\rm{km}\,\rm{s}^{-1}\,\rm{pc}^2) =
3.25\times10^7\,S\Delta v \,\nu_{\rm
  obs}^{-2}\,\rm{D}_{ \rm L}^2\,(1 + z)^{-3},
\end{equation}
where $S\Delta v$ is the integrated flux in units of
$\rm{Jy}\,\rm{km}\,\rm{s}^{-1}$, $\rm{D}_{\rm L}$  the luminosity
distance in Mpc, and $\nu_{\rm obs}$ the observed central frequency of
the line. The integrated flux densities and line and continuum luminosities for 3D-HST GS30274
are presented in Table \ref{tab:submmproperties}.

\subsection{Infrared luminosity, stellar mass, SFR, and dust mass}
We measure a total infrared (IR; 8 - 1000 $\mu$m) luminosity for 3D-HST GS30274 of
$L_{\rm IR} = 2.3\pm0.6\,\times10^{12}\,\rm{L}_\odot$. This was obtained by fitting the ALMA band 4, the \emph{Herschel}/
PACS \citep{Magnelli2013} and SPIRE continuum \citep{Roseboom2010}, and
optical and NIR data photometry \citep{Skelton2014} with the
\texttt{MAGPHYS} \citep[][]{daCunha2008} model (see Figure
  \ref{fig:SED}). 
  These fits suggest a
dust mass of $2.5\pm0.6\,\times10^{8}\,\rm{M}_\odot$, a dust
temperature of  $45\pm 5$ K, and a stellar
mass and SFR of $1.89 \pm 0.47\,\times 10^{11}\,\rm{M}_\odot$ and
$214\pm44\,\rm{M}_\odot\,\rm{yr}^{-1}$, respectively. The inferred IR luminosity
classifies 3D-HST GS30274 as a ULIRG.  This was expected as
  the most massive main-sequence galaxies at $z\sim2$ are indeed ULIRGs
  \citep{Daddi2005, Rodighiero2011}.

Despite the presence of a bright AGN, it is unlikely that it strongly 
affects the SED fitting of 3D-HST GS30274. The
UV-to-sub-mm SED of 3D-HST GS30274 can be well fitted by a stars-only
model. Additionally, the galaxy shows a strong Balmer break at a rest-frame
wavelength of $\lambda_{\rm rest} \sim 4000 \,\AA$, which strongly
constrains the contribution from an AGN to the continuum emission
\citep{Kriek2007,Marsan2015}. Only at observed wavelengths of 8 and 100 $\mu$m does the
\texttt{MAGPHYS} fit underestimate the observed flux. A comparison
between the IR part of the SED of 3D-HST GS30274 and the templates
presented in \citet{Kirkpatrick2015} suggests that an additional contribution of
20\% to the total IR luminosity by an AGN can make up for the mismatch between the
observed and fitted SED.  This contribution would come on top of the
emission already accounted for by stars and star formation.

\subsection{Kinematics}
The spectra presented in Figure \ref{fig:spectra} can be well
fitted with a single gaussian with a dispersion of  328 $\pm$ 38, 321
$\pm$ 19, and 289 $\pm$ 39
km s$^{-1}$ for the CO J$=$3--2, CO J$=$4--3, and [CI] 1--0 lines,
respectively. This corresponds to a FWHM of 764 $\pm$ 88, 751 $\pm$
44, and  675 $\pm$ 92 km
s$^{-1}$, respectively. The
gaussians are represented by the dashed black lines in Figure \ref{fig:spectra}. 
The velocity profile does not show the classic `double horned'
structure that might be expected in a gas disk with a smooth surface
brightness distribution. This may indicate significant random motions,
a complex gas distribution, or a combination. We do not find any signatures in the flux density profile for a CO or [CI] outflow,
such as a broad second gaussian component or a bright CO/[CI]
component blue- and/or redshifted from the source systematic
 velocity. This does not necessarily suggest that no outflow is
 present. If outflows are present the  observations are not deep enough to
retrieve their weak signal. Furthermore, the spatial resolution might
be too low to distinguish a blue- and/or redshifted component away
from the galaxy centre.

\subsection{Comparing line ratios with PDR models}
\label{sec:PDR}
\begin{figure*}[t]
\centering
\includegraphics[width=\linewidth]{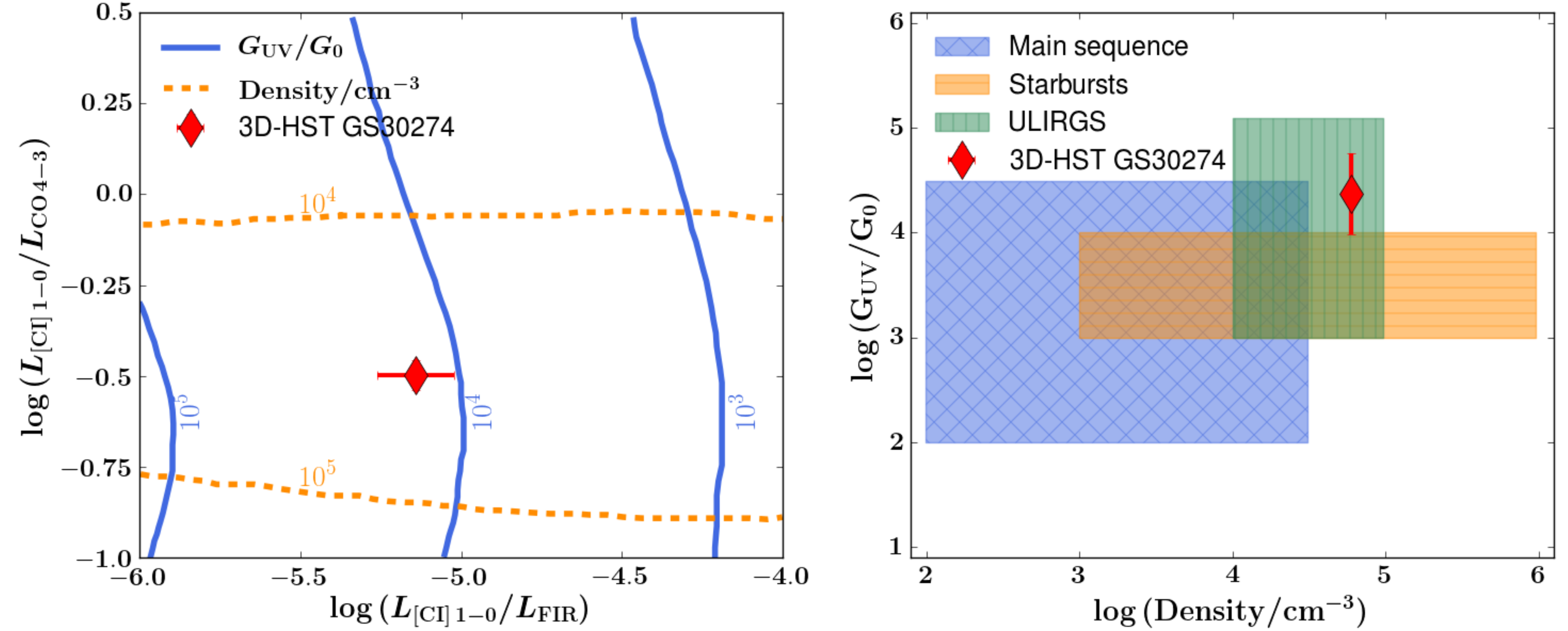}
\centering
 \caption{\label{fig:PDR_plot_Kaufman}{Left: The [CI] 1--0/ CO J$=$4-3 luminosity ratio of 3D-HST
     GS30274 as a function of its [CI] 1--0/ FIR luminosity ratio. In
     this case the luminosities were taken to be in solar
     units. Coloured lines represent the density and UV radiation field
     contours as presented in \citet{Alaghband2013}, based on the
     \citet{Kaufman1999,Kaufman2006} PDR models. These line ratios
     provide us with a good constraint on the UV radiation field and
     add an additional constraint on the gas density. Right: The
     density and UV radiation field of  3D-HST
     GS30274 compared to the densities and radiation fields of
     main-sequence galaxies \citep{Malhotra2001}, local starburst
     galaxies \citep{Stacey1991}, and local ULIRGS
     \citep{Davies2003}. The density of 3D-HST
     GS30274 is similar to densities seen in local starbursts and
     ULIRGS, whereas the UV radiation field is similar to local ULIRGS.}}
\end{figure*}

To explore the properties of the ISM in 3D-HST GS30274 we compare the intensity
ratios of the [CI] and CO lines to the outputs of molecular cloud
models. We present the [CI] 1--0/ CO J$=$4-3 luminosity ratio of 3D-HST
GS30274 as a function of its [CI] 1--0/ FIR luminosity ratio in
Figure \ref{fig:PDR_plot_Kaufman} (left panel). We overplotted the density and UV radiation field
contours as presented in \citet{Alaghband2013}, based on the
\citet{Kaufman1999} and \citet{Kaufman2006} PDR models. With this
approach we find that 3D-HST GS30274 has a density of $(6 \pm 0.5)\times
10^4\,\rm{cm}^{-3}$ and a UV radiation field of $(2 \pm 2) \times
10^4\,\rm{G}_0$. The models of \citet{Meijerink2005} and
\citet{Meijerink2007} suggest that the ISM of 3D-HST GS30274 has a
density of a few times $10^5\,\rm{cm}^{-3}$. Unfortunately, the
Meijerink et al. models do not
constrain the UV radiation field in 3D-HST GS30274 very well.

Although UV-to-sub-mm continuum SED fitting suggest that the AGN does
not dominate the emission from 3D-HST GS30274, we did compare the observed line
intensities to an X-ray Dominated Region (XDR) code
\citep{Meijerink2005,Meijerink2007}. The XDR-code was not able to
reproduce the [CI] 1--0 / CO J$=$3--2 intensity ratio, whereas only
poor convergence was reached for the [CI] 1--0 / CO J$=$4--3 intensity
ratio suggesting densities larger than $10^{6.5}\,\rm{cm}^{-3}$. This
supports the finding that the AGN only plays a minor role in the
emission from 3D-HST GS30274.

 The constraints on the density and UV radiation field of the
  molecular ISM in 3D-HST GS30274 are similar to the conditions
  observed in ultra-luminous infrared galaxies  (ULIRGS) and local (nuclear) starbursts (right panel of
  Figure \ref{fig:PDR_plot_Kaufman}. Multiple independent studies have
  estimated a molecular ISM density of $\sim10^5\,\rm{cm}^{-3}$ in the
  local nuclear starburst NGC 253 \citep{Bayet2004,Meier2015}. The density and UV radiation field in the starburst
region of Arp 220 is estimated to be $10^4$ -- $10^5\,\rm{cm}^{-3}$ and
$\sim4\times10^4 G_0$
\citep[e.g.,][]{Gerin1998,Aalto2009,Scoville2015}. \citet{Israel2015} used the line-ratio between [CI] and CO
to study the ISM in a sample of local (nuclear) starbursts and
estimated an average ISM density of
$10^4$--$10^5\,\rm{cm}^{-3}$. Molecular emission-line studies of the
nuclear region of M82 suggest the presence of dense components with
densities up to $10^6\,\rm{cm}^{-3}$ and a UV radiation field up to
$1.8\times 10^{3}$ \citep{Bayet2008,Loenen2010}. \citet{Pineda2012} found an
average hydrogen density of $\sim 10^5\,\rm{cm}^{-3}$ and a UV
radiation field of $3.1 \times 10^3 G_0$ in the starburst region 30
Doradus. \citet{Chevance2016} estimates a UV radiation field up to
$1.8 \times 10^{4}\,G_0$  in 30 Doradus.  

The estimated ISM properties for 3D-HST
GS30274  are on average in the same range as the ISM properties found
in local ULIRGS and starbursts. This suggests that the molecular ISM in 3D-HST
GS30274 is in a starburst phase.

\subsection{\h2 mass}
In this subsection we will derive the \h2 mass of 3D-HST
GS30274 using three different approaches. We summarise the obtained \h2
masses in Table \ref{tab:H2_masses}.

\subsubsection{Based on [CI]}
To derive the \h2 mass of 3D-HST
GS30274 from the [CI] emission we first
need to estimate the total carbon mass in 3D-HST
GS30274. We derive the atomic carbon mass following
\citet{Weiss2005}, calculated from
the [CI] 1--0 line luminosity as
\begin{equation}
M_{\rm [CI]}(\rm{M}_\odot)= 5.706\times10^{-4}\rm{Q}(T_{\rm ex})\frac{1}{3}e^{23.6/T_{\rm ex}}L'_{[CI](1-0)}\,,
\end{equation}
where $T_{\rm ex}$ is the excitation temperature of [CI] and $\rm{Q}(T_{\rm ex}) = 1 + 3e^{-T_1/T_{\rm ex}} + 5e^{-T_2/T_{\rm ex}}$
  is the [CI] partition function. $T_1$ and $T_2$ are the energies
  above the ground state, 23.6 K and 62.5 K, respectively.

Without knowledge of the [CI] 2--1 luminosity,  we have to make an assumption for the excitation temperature of atomic
carbon. We assume $T_{\rm ex} = 30$ K, the average excitation temperature found in
\citet{Walter2011}. This results in a carbon mass of $5.4 \pm 1.4
\times 10^6\, \rm{M}_\odot$. This carbon mass is
similar to the mass found in $z>2$ dusty SFGs and quasar host
galaxies \citep{Walter2011}. 

The derived density and UV radiation
field would suggest a much higher excitation temperature of $\sim 200$
K for atomic
carbon \citep{Meijerink2007, Kaufman2006}. Here it is important to remember that PDR models assume a
specific geometry and a homogeneous medium (which is probably an
unrealistic assumption). Line intensity ratios are sensitive to  the
chosen geometry, and especially the intensity ratio between [CI] 2--1
and [CI] 1--0 is sensitive to the homogeneity of the medium
\citep{Spaans1996}. This is further demonstrated by the difficulty
models have in reproducing this intensity ratio in environments with
low-UV radiation fields and well defined spherically-symmetric density
profiles \citep[e.g.,][]{Pineda2007,Papadopoulos2004}. \citet{Danielson2011}
observed atomic carbon and CO in a lensed DSFG at $z\sim 2.3$ with
similar density and UV radiation field as 3D-HST
GS30274. Its carbon and
CO line ratios suggest an excitation temperature of a few hundred
K. The carbon excitation temperature directly derived from the
observed [CI] 2--1 and [CI] 1--0 luminosity ratio is 25--45 K, significantly
lower. We therefore stick to the assumed excitation
temperature of  $T_{\rm ex} = 30$ K, rather than adopting high
temperatures of a few hundred degrees. If we were to change the carbon
excitation temperature to for instance the dust temperature of the ISM
in 3D-HST
GS30274 (45 K), this would result in a change in carbon mass (and resulting
\h2 mass) of less than 3\%.

The derived carbon mass of a galaxy can be used to estimate its \h2
mass
\citep[e.g.,][]{Papadopoulos2004,Weiss2005,Wagg2006,Alaghband2013,Bethermin2016,Bothwell2016}. [CI]
is especially a good tracer in X-ray and cosmic-ray dominated
environments \citep{Bisbas2015,Glover2016}. We
estimate the \h2 mass of 3D-HST GS30274 from its atomic carbon mass,
assuming a carbon abundance of  $\rm{X}_{\rm [CI]} = M_{\rm [CI]}/(6\,M_{\rm
  H2}) = 3\times 10^{-5}$
\citep{Weiss2005,Danielson2011,Alaghband2013}. We implicitly make the
assumption that the atomic carbon is mostly neutral. The adopted carbon
abundance is slightly higher than
the Galactic abundance of $2\times 10^{-5}$
\citep{Frerking1989}.   We estimated a
metallicity of $12 + \log{(\rm{O/H})} = 9.07$ for 3D-HST
GS30274 of using the fundamental metallicity relation \citep{mannucci2010}. Our
choice for $\rm{X}_{\rm [CI]} = 3\times 10^{-5}$ agrees well with
theoretical estimates of the carbon abundance in metal-rich
environments by \citet{Glover2016}.  We measure a molecular hydrogen mass of $M_{\rm H2}(\rm{[CI]}) =
3.0 \pm 0.3 \times 10^{10}\,\rm{M}_\odot$. The Galactic abundance of $2\times
10^{-5}$ would have resulted in an \h2 mass of $M_{\rm H2}(\rm{[CI]}) =
4.5 \pm 0.4 \times 10^{10}\,\rm{M}_\odot$. 
 
\subsubsection{Based on CO}
\label{sec:CO}
To use the CO J$=$3--2 emission as a tracer of \h2 mass we first have to derive the CO
J$=$1--0 line luminosity from the CO J$=$3--2 line luminosity. 
The ratio between the squared CO J$=$4--3 and CO J$=$3--2 intensities
is $I_{\rm CO 4-3}^2/I_{\rm CO3-2}^2 = 1.9
\pm 0.3$. This is close to the line ratio one would expect if the
lines are all in the Rayleigh–Jeans limit and in local thermodynamic
equilibrium (LTE; $(\rm{J}=4^2) / (\rm{J} = 3^2) = 1.78$). We therefore assume that
the flux intensity ratio between the CO J$=$3--2 and CO $J$= 1--0 line
equals $3^2/1^2=9$. This gives a line luminosity for CO J$=$1--0 of
$2.1 \pm 0.2 \times 10^9$ K km/s pc$^{2}$. The assumption of a thermalized
medium for the CO transitions is supported by the strong UV radiation
field and high densities derived in Section \ref{sec:PDR}.

In Section \ref{sec:PDR} we concluded that the ISM of 3D-HST
GS30274 resembles the ISM in local starburst and starburst nuclei. We
therefore adopt a CO-to-\h2 conversion factor typically assumed for starburst and mergers \citep[$\alpha_{\rm CO} =
0.8\,\rm{M}_\odot/(\rm{K}\,\rm{km/s}\,\rm{pc}^{2})$,
][]{Downes1998,Bolatto2013}. This leads to a CO-based \h2 mass of 
  $1.7 \pm 0.1 \times 10^{10}\,\rm{M}_\odot$, approximately half of
  the mass derived from [CI].

\subsubsection{Based on the dust mass}
\label{sec:dust}
 An additional method to measure the gas mass of 3D-HST
GS30274 is
to use the inferred dust mass of 3D-HST
GS30274 from the SED fitting in combination with typically assumed gas-to-dust
ratios ($M_{\rm gas} = \delta_{\rm GDR}(Z)M_{\rm dust}$, where
$\delta_{\rm GDR}(Z)$ is the gas-to-dust ratio and a function of
metallicity). We estimate $\delta_{\rm GDR}(Z)$ using the empirical
calibration of \citet{Magdis2012a}. We then find a gas mass (pure hydrogen) of $M_{\rm gas}=
0.7 \pm 0.2 \times 10^{10}\,\rm{M}_\odot$. This mass represents the
combined contribution from atomic and molecular gas and sets an upper
limit on the \h2 mass. We will not attempt to make a correction for
this, though theoretical efforts suggest that a galaxy with similar
stellar mass and SFR as in 3D-HST
GS30274 at
$z\sim 2$ has a molecular hydrogen fraction $M_{\rm H2} / (M_{\rm HI}
+ M_{\rm H2}) \sim 75$\% \citep{Popping2015SHAM}. The derived upper limit
for the \h2 content of 3D-HST GS30274 from the dust mass is already
three times as low as the [CI]-based \h2 estimate.

\begin{table*}[]
\centering
\caption[]{\label{tab:H2_masses}  The \h2 mass, relative amount of \h2, and \h2
  depletion time of 3D-HST
     GS30274 derived by adopting the following approaches; [CI]: an \h2
     mass based on atomic carbon; CO: an \h2 mass based on the CO
     J$=$3--2 luminosity; Dust mass: an \h2 mass based on the inferred dust
   mass from SED fitting in combination with a gas-to-dust ratio.}
\begin{tabular}{cccc}
\hline
\hline
Approach & \h2 mass ($10^{10}\,\rm{M}_\odot)$ & $\rm{M}_{\rm
                                                H2}/(\rm{M}_{\rm H2} +
  \rm{M}_*)$& \h2 depletion time (Myr)\\
\hline 
[CI]&  3.0 $\pm$ 0.3 &0.14 $\pm$ 0.01 & 140 $\pm$30\\
CO starburst $\alpha_{\rm CO}$ & 1.7 $\pm$ 0.1 & 0.08 $\pm$ 0.005& 80
                                                                   $\pm$
  17\\
Dust mass &<0.7 $\pm$ 0.2 & <0.04 $\pm$ 0.01&<30 $\pm$ 10\\
\hline
\end{tabular}
\end{table*}

\subsection{Gas depletion time and galaxy gas fraction}
We can use the measured \h2 masses in combination with the estimated
stellar mass and SFR to calculate the gas fraction and gas depletion
time of 3D-HST
GS30274. The gas fraction that we determine using
the [CI]-based \h2 estimate is $M_{\rm H2} /
(M_{\rm H2} + M_*) =0.14\pm 0.01$. The gas depletion time of 3D-HST
GS30274 is $M_{\rm H2}/{\rm SFR}= 140 \pm 30\,\rm{Myr}$. The gas
fractions obtained using the other approaches are 0.08 $\pm$ 0.005 for
the CO-based \h2 mass and less than 0.04 $\pm$ 0.01, for the
dust-based \h2 mass, whereas the gas depletion times are 80 $\pm$ 17
Myr and less than 30 $\pm$ 10 Myr for the CO-based and dust-based \h2
estimates, respectively. The estimated gas fraction and depletion
times are both low for star-forming galaxies. For example, they are both lower
than what is typically found in the literature for extended star-forming galaxies at similar redshift
and stellar mass \citep{Tacconi2010,Magdis2012a,Tacconi2013,Saintonge2013}. The gas fractions
  found by \citet{Barro2016} for six cSFGs at $z\sim2.5$ are higher than the gas fraction we derived for
  3D-HST GS30274. The gas depletion times on the other hand are similar,
  suggesting 3D-HST GS30274 and the cSFGs in \citet{Barro2016} may have a
  similar mode of star formation.

\subsection{CO-to-\h2 conversion factor}
\label{sec:alphaCO}
The independent measurements of the \h2 mass allow us to constrain the CO-to-\h2 conversion factor in 3D-HST
GS30274, using
the observed CO J$=$3--2 emission line. We first have to derive the CO
J$=$1--0 line luminosity from the CO J$=$3--2 line luminosity, for
which we use the CO J$=$1--0 line luminosity derived in Section
\ref{sec:CO}. We find a
CO-to-\h2 conversion factor of $\alpha_{\rm CO} = \rm{M}_{\rm H2} /
L'_{\rm CO 1-0} = 1.4 \pm 0.15
\,\rm{M}_\odot/(\rm{K}\,\rm{km/s}\,\rm{pc}^{2})$. When using the \h2
masses based on the total dust mass of 3D-HST
GS30274, we find a CO-to-\h2 conversion factor of $0.5
\pm 0.3\,\rm{M}_\odot/(\rm{K}\,\rm{km/s}\,\rm{pc}^{2})$, respectively.

The CO-to-\h2 conversion factors derived for 3D-HST GS30274 lie below the
value typically assumed for main-sequence
galaxies \citep[ranging from 3.6 to 4.3
$\rm{M}_\odot/(\rm{K}\,\rm{km/s}\,\rm{pc}^{2})$, see][for a review]{Bolatto2013}
and around the value usually assumed for starburst and mergers \citep[$\alpha_{\rm CO} =
0.8\,\rm{M}_\odot/(\rm{K}\,\rm{km/s}\,\rm{pc}^{2})$,
][]{Downes1998}. This fits in with the finding that the ISM in 3D-HST
GS30274 is in a starburst phase.

\subsection{The gas-to-dust ratio}
We can constrain the gas-to-dust ratio in 3D-HST
GS30274 using the dust
mass obtained by SED fitting and the \h2 mass obtained from the atomic
carbon line. We find a molecular gas-to-dust ratio of $165\pm 45$ when
using the carbon-based \h2 estimate. Here we included a 1.36
correction for Helium, similar to gas-to-dust ratios quoted in the literature. We emphasise that we have only accounted for
the gas in molecular form and did not correct for atomic
hydrogen. Theoretical models and indirectly estimated gas masses from
observations suggest that the molecular hydrogen fraction $f_{\rm H2}$
for a main-sequence galaxy at $z\sim2$ is around 75 \% \citep[e.g.,]{Popping2015candels}. Taking
this correction into account, the
resulting gas-to-dust ratio is $221\pm60$. When using the CO-based
\h2 estimate instead we find a molecular gas-to-dust ratio of $94 \pm
25$ and a total gas-to-dust ratio of $125\pm33$. The CO-based
gas-to-dust ratio is
similar to the gas-to-dust ratio obtained using the empirical fitting
relation presented in \citet{Magdis2012a}. These observations also confirm that massive
galaxies at $z\sim2$ can already be significantly dust-enriched
\citep[e.g.,][]{Downes1992,Berta2016,Popping2016,Seko2016}, though the exact level of enrichment highly
depends on the chosen \h2 mass estimator.

\section{Discussion}
\label{sec:discussion}
We are now in the unique position to tie the information we obtained
about the cold gas kinematics and physical properties, the gas fraction and the
depletion time, and the studies of ionized gas in the literature together into one complete picture
that aims to describe 3D-HST
GS30274.

The velocity dispersion that we derive for 3D-HST
GS30274 is significantly higher than typically found in extended
star-forming galaxies with similar stellar mass and SFR at the same
redshift \citep[e.g.,][]{Tacconi2010,Tacconi2013}. The derived
velocity dispersions of CO and [CI] are in agreement with the H$\alpha$
velocity dispersion for this source \citep{vanDokkum2015}. \citet{vanDokkum2015} showed
that the observed velocity dispersion of the H$\alpha$ gas in cSFGs 
is consistent with the dispersion expected for a system dominated by
gravitational motions (with little contribution from
outflows). We can repeat this exercise using the [CI] and
CO observations of 3D-HST
GS30274. Following \citet{vanDokkum2015} the predicted
velocity dispersion of a galaxy is
\begin{equation}
\log{\sigma_{\rm pred}} = 0.5 ( \log{M_{*}} - \log{r_{\rm e}} - 5.9).
\end{equation}
Assuming that the gas and stars are dynamically coupled, we can
replace $M_*$ by $M_{\rm dyn} = M_* + 1.36 M_{\rm H2}$ (where the
factor 1.36 accounts for the contribution by Helium). We assume that
the effective radius of 2.5 $\pm$ 0.1 kpc \citep{vanderWel2014} is still valid. When using the [CI]-based \h2 mass, we find a
predicted velocity dispersion for the gas of $340 \pm 35$ km
s$^{-1}$. We predict velocity dispersions of 327 $\pm$ 40
and 316 $\pm$ 40 km s$^{-1}$ for the CO- and dust-based \h2 mass
estimates, respectively. The predicted
velocity dispersions are similar to the observed velocity dispersion
of the CO lines in 3D-HST GS30274 . This suggests that the observed velocity
dispersions are consistent with being gravitationally driven by a compact
distribution of both stars and gas. While AGN or star formation 
driven winds are likely present in this galaxy \citep{Genzel2014}, they do not appear 
to dominate the galaxy-integrated line widths in CO or [CI]. These results are a confirmation that we are likely seeing the build-up of the 
dense core of a massive galaxy. 

The derived ISM characteristics match the
ISM properties observed in local ULIRGS and starbursts very well \citep[e.g.,][see
Section \ref{sec:PDR}]{Bayet2004,Pineda2012,Meier2015}. 
 Additionally, the
derived CO-to-\h2 conversion factor for 3D-HST GS30274 is within the
range typically assumed to starbursting galaxies. We have employed three methods to estimate the \h2 mass of 3D-HST
GS30274. All these approaches yield \h2 depletion times shorter than typically found
in more extended main-sequence galaxies with similar stellar mass and SFR at these
redshifts ($\leq 140$ Myr). Such efficient star-formation fits the picture where a
starburst is quickly and efficiently depleting the gas reservoir of
3D-HST GS30274. 
The short depletion time could also suggest that this galaxy is in 
the process of shutting down its star formation. This is consistent with the picture
in which the star formation in a galaxy is quenched once it has formed 
a high stellar surface density core \citep[e.g.,][]{Kauffmann2003, Franx2008, 
Wuyts2011, Bell2012, Fang2013, vanDokkum2014, 
Lang2014, Barro2015, Whitaker2015,Whitaker2016}.

3D-HST GS30274 is classified as a merger remnant. A likely scenario for this object is that the merger event
triggered compactification of the gas and induced a
central starburst event. This explains the starburst ISM
characteristics, the short depletion time,  and the CO and [CI]
kinematics that are consistent with a compact distribution of the gas.
This is confirmation that we are likely seeing the build-up of the
dense core in 3D-HST GS30274. This scenario is consistent with one of the
scenarios suggested by \citet{Wellons2015} and \citet{Zolotov2015} for the formation of
cSFGs. These models suggest that compaction is triggered by inflow
episodes, such as gas-rich mergers, and is commonly associated with violent
disc instabilities. The high central concentration of gas then quickly
builds up a dense core. It may also accrete
efficiently onto a black hole triggering a bright AGN, as observed in
the X-ray and radio emission from 3D-HST
GS30274. Future high-resolution
observations are needed to properly measure the size of the compact
molecular gas reservoir and better study the kinematic motions. The presented observations demonstrate that the main-sequence of star formation doesn't
solely consist of 'typical star-forming galaxies', but can contain 
galaxies undergoing different modes of star formation.

\section{Summary and Conclusion}
\label{sec:summary}
We have presented ALMA CO, [CI], and dust continuum detections of the
compact star-forming galaxy 3D-HST GS30274. These
  observations mark the first detections of sub-mm emission lines in
  cSFGs and provide a new insight into the origin of cSFGs and how
  they relate to compact quiescent galaxies.

The ratio between the
different sub-mm emission lines suggests that the ISM is in a
starburst phase. This is consistent with the short \h2 depletion time of
$\leq 140 $ Myr we obtained for this source using three different \h2
tracers. The observed CO and [CI] dynamical information for 3D-HST GS30274 is
consistent with a scenario where the gas in 3D-HST GS30274 is situated
in a compact reservoir.

The presented observations point towards a scenario where a previous merger
has driven the gas in 3D-HST GS30274 towards a compact distribution in
the galaxy centre and triggered a nuclear starburst event. The
starburst event quickly depletes the cold gas and has started to build up a
dense stellar core. The compact distribution of gas may feed a
black hole which triggers the observed AGN. Ultimately, the
combination of short depletion times, the AGN activity, and possibly
star-formation driven outflows may quench 3D-HST GS30274 into a compact
quiescent galaxy.

The presented work on 3D-HST GS30274 demonstrates that ALMA detections
of sub-mm lines to probe the molecular gas mass, ISM characteristics,
and molecular gas dynamics of cSFGs are an important puzzle piece to understand the formation
history of the progenitors of compact quiescent galaxies.

\begin{acknowledgements}
A large part of the analysis and writing of this paper was carried out at the
ESO guesthouse in Santiago. GP ('the appendicitis guy') thanks the
staff of the ESO guesthouse for its hospitality and great
care. We thank the referee for a very constructive report. GP thanks Karina Caputi, Juan Pablo P\'erez-Beaupuits, and Rachel Somerville for stimulating discussions during the first stages of this project. Support for RD
was provided by the DFG priority program 1573 `The physics of the
interstellar medium'. BG acknowledges support from the ERC Advanced
Investigator programme DUSTYGAL 321334. This paper makes use of the following ALMA data:
  ADS/JAO.ALMA\#2015.1.00228.S. ALMA is a partnership of ESO (representing its member states), 
  NSF (USA) and NINS (Japan), together with NRC (Canada), NSC and ASIAA (Taiwan), and KASI 
  (Republic of Korea), in cooperation with the Republic of Chile. 
  The Joint ALMA Observatory is operated by ESO, AUI/NRAO and NAOJ.
\end{acknowledgements}

\bibliographystyle{aa} 
\bibliography{references} 
\begin{appendix}

\newpage
\section{The photometry for 3D-HST GS30274 adopted in the SED fitting}
\begin{table}[!h]
\centering
\caption{\label{tab:photometry}
The photometry of 3D-HST GS30274 used for the SED fitting.
The fluxes in this table were compiled from the following references: (1) \citet[and references therein]{Skelton2014}, (2)
  \citet{Magnelli2013}, (3) \citet{Roseboom2010}.}
\begin{tabular}{ccc}
\hline \noalign {\smallskip}
\hline \noalign {\smallskip}
Wavelength & Flux & Reference\\
$\mu$m & Jy & \\
\hline \noalign {\smallskip}
0.3676 & 3.83$\times10^{-7}$ $\pm$ 4.39$\times10^{-8}$ & 1 \\
0.4263 & 8.54$\times10^{-7}$ $\pm$ 5.88$\times10^{-8}$ & 1 \\
0.4328 & 8.12$\times10^{-7}$ $\pm$ 1.20$\times10^{-8}$ & 1 \\
0.4603 & 8.36$\times10^{-7}$ $\pm$ 1.60$\times10^{-8}$ & 1 \\
0.5063 & 9.83$\times10^{-7}$ $\pm$ 4.91$\times10^{-8}$ & 1 \\
0.5261 & 1.11$\times10^{-6}$ $\pm$ 2.56$\times10^{-8}$ & 1 \\
0.5378 & 1.09$\times10^{-6}$ $\pm$ 2.04$\times10^{-8}$ & 1 \\
0.5764 & 1.41$\times10^{-6}$ $\pm$ 5.53$\times10^{-8}$ & 1 \\
0.5959 & 1.39$\times10^{-6}$ $\pm$ 9.73$\times10^{-9}$ & 1 \\
0.6233 & 1.55$\times10^{-6}$ $\pm$ 2.75$\times10^{-8}$ & 1 \\
0.6781 & 1.61$\times10^{-6}$ $\pm$ 2.39$\times10^{-8}$ & 1 \\
0.7361 & 1.93$\times10^{-6}$ $\pm$ 3.14$\times10^{-8}$ & 1 \\
0.7684 & 2.09$\times10^{-6}$ $\pm$ 7.37$\times10^{-8}$ & 1 \\
0.7705 & 2.11$\times10^{-6}$ $\pm$ 1.59$\times10^{-8}$ & 1 \\
0.9048 & 2.76$\times10^{-6}$ $\pm$ 2.15$\times10^{-8}$ & 1 \\
0.914 & 2.44$\times10^{-6}$ $\pm$ 1.17$\times10^{-7}$ & 1 \\
1.2369 & 5.65$\times10^{-6}$ $\pm$ 1.08$\times10^{-7}$ & 1 \\
1.2499 & 6.86$\times10^{-6}$ $\pm$ 9.88$\times10^{-8}$ & 1 \\
1.5419 & 1.09$\times10^{-5}$ $\pm$ 4.80$\times10^{-8}$ & 1 \\
1.6464 & 1.17$\times10^{-5}$ $\pm$ 1.87$\times10^{-7}$ & 1 \\
2.1454 & 1.90$\times10^{-5}$ $\pm$ 1.62$\times10^{-7}$ & 1 \\
2.2109 & 1.93$\times10^{-5}$ $\pm$ 1.40$\times10^{-7}$ & 1 \\
3.6 & 2.52$\times10^{-5}$ $\pm$ 1.20$\times10^{-7}$ & 2 \\
4.5 & 3.39$\times10^{-5}$ $\pm$ 1.33$\times10^{-7}$ & 2 \\
5.8 & 4.83$\times10^{-5}$ $\pm$ 5.91$\times10^{-7}$ & 2 \\
8.0 & 6.72$\times10^{-5}$ $\pm$ 6.66$\times10^{-7}$ & 2 \\
24 & 3.57$\times10^{-4 }$$\pm$ 1.01$\times10^{-5}$ & 2 \\
100 & 3.53$\times10^{-3}$ $\pm$ 2.45$\times10^{-4}$ & 2 \\
170 & 7.30$\times10^{-3}$ $\pm$ 7.03$\times10^{-4}$ & 2 \\
250 & 1.19$\times10^{-2}$ $\pm$ 2.62$\times10^{-3}$ & 3 \\
350 & 1.01$\times10^{-2}$ $\pm$ 3.19$\times10^{-3}$ & 3 \\
500 & 3.19$\times10^{-3}$ $\pm$ 4.27$\times10^{-3}$ & 3 \\
2200 & 8.5$\times10^{-5}$ $\pm$ 1.4$\times10^{-5}$ & This work \\
\hline \noalign {\smallskip}
\end{tabular}
\end{table}
\end{appendix}
\end{document}